# Data Quality Principles in the Semantic Web

Ahmad Assaf and Aline Senart

SAP Research, SAP Labs France SAS

805 avenue du Dr. Maurice Donat, BP 1216, 06254 Mougins Cedex, France

firstname.lastname@sap.com

*Abstract*— The increasing size and availability of web data make data quality a core challenge in many applications. Principles of data quality are recognized as essential to ensure that data fit for their intended use in operations, decision-making, and planning. However, with the rise of the Semantic Web, new data quality issues appear and require deeper consideration. In this paper, we propose to extend the data quality principles to the context of Semantic Web. Based on our extensive industrial experience in data integration, we identify five main classes suited for data quality in Semantic Web. For each class, we list the principles that are involved at all stages of the data management process. Following these principles will provide a sound basis for better decision-making within organizations and will maximize long-term data integration and interoperability.

## I. INTRODUCTION

Data quality is complex and involves data management, modeling, analysis, storage and presentation, quality control and assurance [1]. Moreover, data quality is subjective and as the saying goes "beauty is in the eye of the beholder". Data quality cannot indeed be assessed easily and independently by the user. The actual value of data is realized when it is used [2], thus the quality relates directly to the ability of satisfying the users' continuous needs. It was found out that many data quality problems are in fact "data misinterpretations", or problems with the data semantics [3]. For example, the P/E ratio [1] obtained for a certain stock from several financial information systems can be different. The ambiguity is caused by the fact that each source can have its own interpretation and application of the financial term P/E; the earnings in one source can be for one year where it is only defined quarterly in another. Moreover, different sources having the earnings defined for one year can have different interpretations for it; is it the calendar year, fiscal year, or the last 12 months?

The rise of Semantic Web in recent years was followed by a tremendous increase in the amount of data; everyone today has the ability to publish and retrieve information to be consumed or integrated into their applications. The Semantic Web has significantly changed people's perceptions of the Internet. The Semantic Web is seen as a "global database" [5] that machines can directly access and naturally understand [6]. Lots of organizations are therefore trying to leverage external data sources from the Semantic Web like social media feeds, weblogs, sensor data or data published by governments or organizations [7] in order to produce more informed business decisions.

However, these external sources exhibit heterogeneous models, formats and terminologies. Finding and retrieving accurate information on demand is very difficult for organizations. Lots of work is being done to improve the quality of this structured knowledge [8][9] but data quality for Semantic Web is mainly performed in silos without following general methodologies. Tools and best practices are therefore required to help data consumers identify their needs and evaluate the quality of data.

Based on our extensive experience in data integration at SAP, we identify in this paper five principle classes to describe the quality of a particular linked dataset. For each class, we list the principles that are involved at all stages of the data management process. Using our principles, it becomes possible to automate the process of controlling and guaranteeing data quality and consequently to increase the quality of decisions in a business environment.

The remainder of this paper is organized as follows: Section II presents the related work; Section III defines the classification of the principles; and finally Section IV presents concluding remarks and identifies promising areas of research.

## II. RELATED WORK

Semantic data is widely available and the development and application of ontologies have been gaining big momentum in a range of application domains, such as government organizations, healthcare or media [10][11][12]. Several semantic groups have been building or contributing to the development of ontologies, for example [13]. Although numerous methodologies and design patterns exist to support their building process and ensure better data quality, ontologies exhibit heterogeneous structure and content. Deciding what ontology to use becomes one of the most difficult and challenging task for organizations.

Some projects have proposed solutions to identify good data sources simplifying greatly the task of finding and consuming high-quality data. In [14][15] a resource is ranked by the quality of the incoming and outgoing links. Moreover, "Sieve" [16] is a framework that tries to express quality assessment methods as well as fusion methods.

---

[1] A measure of the price paid for a share relative to the annual Earnings per Share [4]

Although these projects go in the right direction, there is still a need for data quality principles in the Semantic Web context. An initial attempt to identify quality criteria for Linked Data sources can be found in [22]. Though this classification is good, some criteria on the quality of the used ontologies and the links between data and ontology concepts are missing. In this paper, we extend this initial list by considering all possible criteria from multiple context factors and by selecting the most relevant indicators to assess data quality in the Semantic Web.

## III. CLASSIFICATION OF DATA QUALITY PRINCIPLES IN THE SEMANTIC WEB

The main goal behind using Linked Data is to easily enable knowledge sharing and publishing. The basic assumption is that the usefulness of Linked data will increase if it is more interlinked with other data; Tim Berners-Lee defined 4 keys principles for publishing [17]:

- Make the data available on the web: assign URIs to identify things.
- Make the data machine readable: use HTTP URIs so that looking up these names is easy.
- Use publishing standards: when the lookup is done provide useful information using standards like RDF.
- Link your data: include links to other resources to enable users to discover more things.

By following these guidelines, a certain level of uniformity is achieved, which increases the usability of data. To fully leverage all the benefits of the Semantic Web, data quality principles in Semantic Web should embrace and adopt the guidelines for Linked Open Data.

Building on these principles and based on our experience with powerful data integration software to extract, transform, and load data from applications, databases and other data sources, we have derived five principles for data quality in the Semantic Web (see Table 1). These principles are:

- **Quality of data source**: This principle is related to the availability of the data and the credibility of the data source.
- **Quality of raw data**: This principle is mainly related to the absence of duplicates, entry mistakes, and noise in the data.
- **Quality of the semantic conversion**: This principle is related to the transformation of raw data into rich data by using vocabularies.
- **Quality of the linking process**: This principle is related to the quality of links between two datasets.

- **Global quality**: This principle is cross-cutting the other principles and covers the source, raw data, semantic conversion, reasoning and links quality.

| Data Quality Principle | Attribute | |
|---|---|---|
| Quality of Data Sources | Accessibility | |
| | Authority & Sustainability | |
| | License | |
| | Trustworthiness & verifiability | |
| | Performance | |
| Quality of raw data | Accuracy | Referential correspondence |
| | | Cleanness |
| | | Consistency |
| | Comprehensibility | |
| | Completeness | |
| | Typing | |
| | Provenance | |
| | Versatility | |
| | Traceability | |
| Quality of the semantic conversion | Correctness | |
| | Granularity | |
| | Consistency | |
| Quality of the linking process | Connectedness | |
| | Isomorphism | |
| | Directionality | |

Table 1- Data quality principles in the Semantic Web

### A. Quality of data source

This principle is related to the availability of the data and the credibility of the data source.

- **Accessibility**: Do access methods and protocols perform properly? Is all the URIs de-referenceable? Do the in-going and out-going links operate correctly?
- **Authority & Sustainability**: Is the data source provider a known credible source or is he sponsored by well-known associations and providers? Are there credible basis for believing the data source will be maintained and available in the future?
- **License**: Is the data source license clearly defined?
- **Trustworthiness & Verifiability**: Can the data consumer examine the correctness and accuracy of the data source? The consumer should also be sure that the data he receives is the same data he has vouched for and from the same resource. This can be ensured using digital signatures thus verifying all possible serialization of that data [18].
- **Performance**: Is the data source capable of coping with increasing requests in low latency response time and high throughput?

### B. Quality of the raw data

This principle is mainly related to the absence of duplicates, entry mistakes, and noise in the data.

- **Accuracy**: Are the nodes referring to factually and lexically correct information?
    - **Referential correspondence**: Is the data described using accurate labels without duplications? The goal is to have one-to-one references between data and real world.
    - **Cleanness**: Is the data clean and not polluted with irrelevant or outdated data? Are there duplicates? Is the data formatted in a consistent way (i.e., are the dates all formatted yyyy/mm/dd)? Tools such as Google Refine [19] or Data Wrangler [20] provide already a good answer to these issues by allowing the cleaning of complex data sets.
    - **Consistency**: does the data contradict itself? For example, is the population of Europe the same as the sum of the population of the European countries? To achieve that we need to validate the underlying vocabulary and syntax of the document with other resources
- **Comprehensibility**: Are the data concepts understandable to humans? Do they convey logical meaning of the described entity and allow easy consumption and utilization of the data? If a concept is described using multiple labels (a set of concepts in a owl:sameAs relationship), which one should be consumed? How can we specify which label is canonical?
- **Completeness**: Do we have all the data needed to represent all the information related to a real world entity? Moreover, is the data related or linked to this set complete as well, e.g., all European countries, all French cities, all street addresses, all postal codes…?
- **Typing**: Is the data properly typed as a concept from a vocabulary or just as a string literal? Having the data properly typed allows users to go a step further in the business analysis and decision process.
- **Provenance**: provenance in the Semantic Web is considered as one of the most important indicators of "quality." Data sets can be used or rejected depending on the availability of sufficient and/or relevant metadata attached.
- **Versatility**: Can the data provided be presented using alternative representations? This can be achieved by conversion into various formats or if the data source enables content negotiation.
- **Traceability**: Are all the elements of my data traceable (including data itself but also queries, formulae)? Can I know from what data sources they come?

*C. Quality of the semantic conversion*

Semantic conversion is the process of transforming "normal" raw data into "rich" data, i.e. input: [tabular data] → output: [RDF using x Vocabulary]. The use of high quality vocabularies and the efficiency of data discovery process are major factors in increasing the quality of data. However, one of the most important aspect that affects the quality of the semantic conversion is the quality and suitability of its data model with the intended usage. The quality of a data model strongly depends on the following aspects:

- **Correctness**: Is the data structure properly modeled and presented for future conversion?
- **Granularity**: does the model capture enough information to be useful? Are all the expected data present?
- **Consistency**: Is the direction of relations consistently done?

*D. Moreover, there shouldn't be any redefinitions of existing properties and no stating of inconsistent values for them. Quality of the linkage*

This principle is related to the quality of links between two datasets.

- **Connectedness**: Is the combination of datasets done at the correct resources? Frameworks like Silk [21] ease the linking process but don't tackle per se the quality of the links that are generated. The quality depends on the link generation configuration. The quality is however improved if your data is linked to some reference dataset.
- **Isomorphism**: Are the combined datasets modeled in a compatible way? Are the combined models reconciled?
- **Directionality**: After the linkage, is the knowledge represented in the resulting graph of resources still consistent?

*E. Global quality*

These principles are applicable to all aspects of a Semantic System (data source, raw data, links, etc.).

- **Timeliness**: Is the data up-to date? Does the data source contain the latest raw data presented with the last updated model? Are the links from and to the data source updated to the latest references? Does the source state the update and validation frequencies? Failing in updating the source data increases the chance that the referenced URIs have changed.
- **History**: Can we keep track of who edited my data and when?
- **Freshness**: The ability to replicate the remote repository into local triple stores and maintain the timeliness of the replica.

IV. SUMMARY

In this paper, we presented five main classes of data quality principles for the Semantic Web. For each class, we listed the specific criteria that represent the quality of a data source on the Web. This new vocabulary to express Linked Data quality can be used by data publishers to refine and improve their datasets, and by consumers to select the most relevant public datasets with highest quality. Following these principles will lead to higher quality Semantic Web, which will result in better data usage and mash-ups thus more informed decisions.

Trust issues have always been dominant in the world of the Internet, no one believes everything that is out there, but rather relies on context, provenance and authority. If the data source cannot be directly trusted then users generally question the data. The unique problem for Linked data is that it considers data as a big graph that originates from users all over the world. The borders of provenance in this case can easily become vague especially when trying to infer across multiple datasets. We will investigate these issues in future work.